\documentclass[12pt]{article}%
\usepackage{amssymb}%
\usepackage{amsmath}%
\usepackage{amsfonts}%
\usepackage{graphicx}

\begin{document}

\title{Modulated phases and devil%
\'{}%
s staircases in a layered mean-field version of the ANNNI model}
\author{Eduardo Nascimento$^{(1,2)}$, J. P. de Lima$^{(1)}$, and S. R. Salinas$^{(2)}$\\(1) 
Departamento de F\'{\i}sica, Universidade Federal do Piau\'{\i},\\CEP 64049--550, Teresina, PI, Brazil\\(2) 
Instituto de F\'{\i}sica, Universidade de S\~{a}o Paulo,\\Caixa Postal 66318, CEP 05314-970, S\~{a}o Paulo, SP, Brazil}
\maketitle

\begin{abstract}
We investigate the phase diagram of a spin-$1/2$ Ising model on a cubic
lattice, with competing interactions between nearest and next-nearest
neighbors along an axial direction, and fully connected spins on the sites of
each perpendicular layer. The problem is formulated in terms of a set of
noninteracting Ising chains in a position-dependent field. At low
temperatures, as in the standard mean-feild version of the
Axial-Next-Nearest-Neighbor Ising (ANNNI) model, there are many distinct
spatially commensurate phases that spring from a multiphase point of
infinitely degenerate ground states. As temperature increases, we confirm the
existence of a branching mechanism associated with the onset of higher-order
commensurate phases. We check that the ferromagnetic phase undergoes a
first-order transition to the modulated phases. Depending on a parameter of
competition, the wave number of the striped patterns locks in rational values,
giving rise to a devil%
\'{}%
s staircase. We numerically calculate the Hausdorff dimension $D_{0}$
associated with these fractal structures, and show that $D_{0}$ increases with
temperature but seems to reach a limiting value smaller than $D_{0}=1$.

\end{abstract}

\section{Introduction}

The Axial-Next-Nearest-Neighbor Ising (ANNNI) model, which includes competing
ferro and antiferromagnetic interactions between pairs of spins along an axial
direction, is known to display a spectacularly rich phase diagram, with a host
of modulated phases \cite{bak88}\cite{selke88}\cite{yeomans88}\cite{selke92}.
The ANNNI model is perhaps the simplest lattice statistical model to account
for the presence of spatially modulated phases in a large variety of physical
systems \cite{ohwada07}\cite{ouisse12}.

The energy of the ANNNI model on a simple cubic lattice may be written as%
\begin{equation}
\mathcal{H}_{ANNNI}=-\frac{1}{2}\sum_{x,y,z}\left[  J_{0}\sigma_{x,y,z}%
\sigma_{x\pm1,y\pm1,z}+J_{1}\sigma_{x,y,z}\sigma_{x,y,z\pm1}+J_{2}%
\sigma_{x,y,z}\sigma_{x,y,z\pm2}\right]  ,\label{annni1}%
\end{equation}
where the sum is over all lattice sites, and $\sigma_{x,y,z}=\pm1$ is a
spin-$1/2$ variable at site $(x,y,z)$. We assume ferromagnetic interactions,
$J_{0}>0$, between nearest-neighbor sites on the $x-y$ planes, and competing
ferromagnetic, $J_{1}>0$, and antiferromagnetic, $J_{2}<0$, interactions
between nearest and next-nearest neighbors along the axial $z$ direction. We
then introduce a parameter $p=-J_{2}/J_{1}>0$ to gauge the strength of the
competitions, and look at the $T-p$ phase diagram, where $T$ is the absolute
temperature. At zero temperature, with $p<1/2$, one easily shows that the
ground state is a trivial ferromagnet. For $p>1/2$, however, the ground state
displays a peculiar antiferromagnetic structure, which has been called a
$\left\langle 2\right\rangle $ phase, with two planes of $+$ spins followed by
two planes of $-$ spins, along the $z$ direction. In the special (multiphase)
point $p=1/2$, the ground state becomes infinitely degenerate, with the
coexistence of a ferromagnetic phase, the antiphase $\left\langle
2\right\rangle $, associated with a period of $4$ lattice spacings along the
$z$ direction, and a multiplicity of modulated phases of larger periods
\cite{selkef79}\cite{szpilkaf86}.

Several theoretical approaches have been used to account for the complex $T-p
$ phase diagram of the ANNNI model, including careful layer-by-layer
mean-field calculations \cite{bakb80}\cite{yokoi81}\cite{selked84}, Monte
Carlo simulations \cite{selkef79}\cite{kaskis85} and analyses of (exact)
low-temperature series expansions \cite{selkef79}\cite{szpilkaf86}. At finite
temperature, all of these calculations indicate the springing from the
multiphase point of larger-period modulated phases. In particular, early
mean-field calculations by Selke and Duxbury \cite{selked84}, which are in
asymptotic agreement with the analysis of the low-temperature series
expansions, support the existence of a branching process of ramification that
explains the onset of new modulated phases at higher temperatures. More recent
self-consistent \cite{surda04}\cite{gendiarn05} and Monte Carlo
\cite{zhangc10} calculations may differ in a number of details, but do confirm
the general qualitative features of the $T-p$ phase diagrams.

Taking into account the relevance of the ANNNI model, and some remaining
questions about the $T-p$ phase diagram in the region of intermediate
temperatures, in special the need of a better characterization of the devil%
\'{}%
s staircase behavior of the succession of commensurate modulated structures,
we decided to revisit this problem and check some points. We then consider the
exact formulation of a special ANNNI model, with fully connected spins at each
layer, which amounts to solving the original problem in the layer-by-layer
mean-field approximation with the addition of spin fluctuations along the
axial direction. In other words, we investigate the effects of the
introduction of additional fluctuations in the old mean-field calculations.
This layered-connected ANNNI model, which we call LC-ANNNI, can also be
obtained from the usual ANNNI model Hamiltonian on a hypercubic lattice in the
limit of infinite coordination of the spins on each layer. In this limit, we
assume a coordination $q_{\perp}\longrightarrow\infty$ within each layer, with
pair interactions of the form $J_{0}/q_{\perp}$, and a fixed value of $J_{0}$.
Spin variables on each layer are fully connected, but we preserve the
short-range character of the competing interactions (and correlations) along
the $z$ direction. The free energy of this special model can be written
exactly, leading to equations of state that can be numerically analyzed in
great detail.

The layout of this paper is as follows. In Section II we define the LC-ANNNI
model, write an exact expression for the free energy, and establish the
equations of state, which are amenable to a detailed numerical analysis. Also,
we describe the equivalent layer-by-layer mean-field approximation for the
analogous ANNNI model. A global $T-p$ phase diagram is obtained in Section
III, which also contains some comments on the previous mean-field results.
Except for some expected quantitative changes, due to taking into account
additional axial fluctuations, we do agree with the mean-field calculations of
Selke and Duxbury \cite{selked84}, including the branching process, and the
recovery of the asymptotic domain-wall analysis near the multiphase point. We
draw some graphs of the main wave number of the modulated structures as a
function of $p$, for fixed values of $T$, and perform a detailed numerical
analysis of the fractal character of these devil%
\'{}%
s staircases. We obtain numerical values for the Hausdorff dimension $D_{0}<1
$ of these fractal structures, and show that $D_{0}$ increases with
temperature, with a limiting value $D_{0}\approx0.8$, which seems to be a
common feature of several problems represented by area-preserving maps
\cite{biham89}, and does support the view that the commensurate modulated
structures occupy most of the ordered region of the $T-p$ phase diagram. Some
concluding remarks are presented in the final Section.

\section{The LC-ANNNI model}

The Hamiltonian of the analog of the ANNNI model with fully connected spins at
each layer, which we call LC-ANNNI model, is given by a sum of a long-range,
mean-field term, $\mathcal{H}_{lr}$, and a short-range term that includes the
axial interactions,%
\begin{equation}
\mathcal{H}=\mathcal{H}_{lr}+\mathcal{H}_{sr},\label{annni2}%
\end{equation}
with%
\begin{equation}
\mathcal{H}_{lr}=-\sum_{z=1}^{N}\frac{J_{0}}{2N^{2}}\left(  \sum_{x,y}%
\sigma_{x,y,z}\right)  ^{2},\label{lr}%
\end{equation}
and%
\begin{equation}
\mathcal{H}_{sr}=-\sum_{x,y,z}\left[  J_{1}\sigma_{x,y,z}\sigma_{x,y,z+1}%
+J_{2}\sigma_{x,y,z}\sigma_{x,y,z+2}\right]  ,\label{sr}%
\end{equation}
where we assume that there are $N$ sites along the sides of a cubic lattice.
It should be remarked that $\mathcal{H}_{sr}$ includes the axial short-range
interactions, and $\mathcal{H}_{lr}$ represents the long-range, mean-field,
ferromagnetic interactions between all pairs of sites on each plane
perpendicular to the $z$ direction. We then write the partition function,
\begin{equation}
Z=\sum_{\left\{  \sigma_{x,y,z}\right\}  }\exp\left[  \sum_{z}\frac{\beta
J_{0}}{2N^{2}}\left(  \sum_{x,y}\sigma_{x,y,z}\right)  ^{2}-\beta
\mathcal{H}_{sr}\right]  ,
\end{equation}
where $\beta=1/k_{B}T$ and the first sum is over all spin configurations.

Using a set of Gaussian identities,%
\begin{equation}
\int_{-\infty}^{+\infty}\exp\left(  -x^{2}+2ax\right)  \,\frac{dx}{\sqrt{\pi}%
}=\exp\left(  a^{2}\right)  ,
\end{equation}
and discarding some irrelevant terms, it is straightforward to write the more
convenient expression%
\begin{equation}
Z=\int dm_{1}\dotsb\int dm_{N}\exp\left(  -\beta N^{3}\phi\right)  ,
\end{equation}
where%
\begin{equation}
\phi=\frac{J_{0}}{2N}\sum_{z=1}^{N}m_{z}^{2}-\frac{1}{\beta N}\ln Z_{I}\left(
\left\{  m_{z}\right\}  \right)  ,\label{fi0}%
\end{equation}
and $Z_{I}\left(  \left\{  m_{z}\right\}  \right)  $ is the partition function
of an Ising chain (with competing interactions) in the presence of
site-dependent effective fields, $\lbrace J_{0}m_{z}\rbrace$,
\begin{equation}
Z_{I}\left(  \left\{  m_{z}\right\}  \right)  =\sum_{\left\{  \sigma
_{z}\right\}  }\exp\left[  \sum_{z=1}^{N}\left(  \beta J_{0}m_{z}\sigma
_{z}+\beta J_{1}\sigma_{z}\sigma_{z+1}+\beta J_{2}\sigma_{z}\sigma
_{z+2}\right)  \right]  ,
\end{equation}
where $\left\{  \sigma_{z}=\pm1\right\}  $, for $z=1$, $2$, ..., $N$, is a
short-hand notation for the spin variables $\left\{  \sigma_{x,y,z}%
=\pm1\right\}  $.

To perform calculations in the ordered regions of the phase diagram, it is
convenient to use a transfer matrix technique and write%
\begin{equation}
\phi=\frac{J_{0}}{2N}\sum_{z=1}^{N}m_{z}^{2}-\frac{1}{\beta N}\ln\left(
\operatorname*{Tr}\prod_{z=1}^{N}\mathbf{V}_{z}\right)  ,\label{fi}%
\end{equation}
where the $4\times4$ matrix $\mathbf{V}_{z}$ is given by
\begin{align}
\mathbf{V}_{z}  & =\left(
\begin{array}
[c]{cccc}%
y_{z} & 0 & 0 & 0\\
0 & y_{z} & 0 & 0\\
0 & 0 & y_{z}^{-1} & 0\\
0 & 0 & 0 & y_{z}^{-1}%
\end{array}
\right)  {}\nonumber\\
& \times\left(
\begin{array}
[c]{cccc}%
x_{1}x_{2} & x_{2}^{-1} & 0 & 0\\
0 & 0 & x_{2}^{-1} & x_{1}^{-1}x_{2}\\
0 & 0 & x_{1}x_{2} & x_{2}^{-1}\\
x_{2}^{-1} & x_{1}^{-1}x_{2} & 0 & 0
\end{array}
\right)  ,
\end{align}
with%

\begin{equation}
x_{1}=\exp(\beta J_{1}),\quad x_{2}=\exp(\beta J_{2}),\quad y_{z}=\exp(\beta
J_{0}m_{z}).
\end{equation}
In the thermodynamic limit, $N\rightarrow\infty$, the asymptotic form of the
partition function comes from an application of Laplace%
\'{}%
s method.

Given a commensurate phase, the magnetization profile is repeated after a
certain finite number $n$ of layers. Therefore, without any loss of
generality, we consider $N=nM$ in Eq. (\ref{fi}). In the thermodynamic limit,
we then write the free energy functional%
\begin{equation}
\phi=\frac{J_{0}}{2n}\sum_{z=1}^{n}m_{z}^{2}-\frac{1}{\beta n}\ln\lambda
_{0}\,,\label{fi2}%
\end{equation}
where $\lambda_{0}$ is the maximum eigenvalue of the matrix
\begin{equation}
\mathbf{V}=\prod_{z=1}^{n}\mathbf{V}_{z}\,.
\end{equation}
The equilibrium magnetization pattern, $\{m_{z}\}$, comes from the stationary
conditions, $\partial\phi/\partial m_{z}=0$, which lead to a set of nonlinear
coupled equations,
\begin{equation}
m_{z}=\frac{\langle l_{0}|\mathbf{V}_{1}\mathbf{V}_{2}\cdots\mathbf{V}%
_{z-1}\mathbf{S}\mathbf{V}_{z}\mathbf{V}_{z+1}\cdots\mathbf{V}_{n}%
|r_{0}\rangle}{\lambda_{0}\langle l_{0}|r_{0}\rangle}\,,\label{eme}%
\end{equation}
where $|r_{0}\rangle$ and $|l_{0}\rangle$ are the right and left eigenvectors
of the transfer matrix $\mathbf{V}$, corresponding to $\lambda_{0}$, and the
matrix $\mathbf{S}$ is given by
\begin{equation}
\mathbf{S}=\left(
\begin{array}
[c]{cccc}%
1 & 0 & 0 & 0\\
0 & 1 & 0 & 0\\
0 & 0 & -1 & 0\\
0 & 0 & 0 & -1
\end{array}
\right)  .
\end{equation}
At fixed values of $T$ and $p$, the stable magnetization profile comes from
the solution of Eq. (\ref{eme}) that minimizes the free energy functional
$\phi$, given by Eq. (\ref{fi2}).

\subsection{Equivalent mean-field approximation}

Consider the Hamiltonian of the ANNNI model on a cubic lattice, given by Eq.
(\ref{annni1}). A mean-field solution for this problem can be obtained from
the variational inequality%
\begin{equation}
G\left(  \mathcal{H}\right)  \leq G_{0}\left(  \mathcal{H}_{0}\right)
+\left\langle \mathcal{H}-\mathcal{H}_{0}\right\rangle _{0}=\Phi,
\end{equation}
where $G\left(  \mathcal{H}\right)  $ is the free energy of the system,
$G_{0}\left(  \mathcal{H}_{0}\right)  $ is the free energy of a system
associated with a trial Hamiltonian $\mathcal{H}_{0}$, and $\left\langle
...\right\rangle _{0}$ is an average value with respect to $\mathcal{H}_{0}$.
In the usual layer-by-layer mean-field calculations \cite{yokoi81}, we use a
free trial Hamiltonian,%
\begin{equation}
\mathcal{H}_{0}=-%
{\displaystyle\sum\limits_{x,y,z}}
\eta_{z}\sigma_{x,y,z},
\end{equation}
where $\left\{  \eta_{z}\right\}  $ is a set of field (variational)
parameters. It is easy to write $\Phi=\Phi\left(  \left\{  \eta_{z}\right\}
\right)  $, and obtain the mean-field solutions by minimizing this expression
of $\Phi$ with respect to the field parameters.

In order to include fluctuations along the $z$ direction, we consider another
trial Hamiltonian,
\begin{equation}
\mathcal{H}_{01}=-\frac{1}{2}%
{\displaystyle\sum\limits_{x,y,z}}
\left[  J_{1}\sigma_{x,y,z}\sigma_{x,y,z\pm1}+J_{2}\sigma_{x,y,z}%
\sigma_{x,y,z\pm2}\right]  -%
{\displaystyle\sum\limits_{x,y,z}}
\eta_{z}\sigma_{x,y,z},
\end{equation}
which corresponds to independent Ising chains along the $z$ direction. For a
cubic lattice, with $N\times N\times N$ sites, it is easy to show that
\begin{equation}
G_{01}=-\frac{N^{2}}{\beta}\ln Z_{01},
\end{equation}
with%
\begin{equation}
Z_{01}=%
{\displaystyle\sum\limits_{\left\{  \sigma_{z}\right\}  }}
\exp\left[
{\displaystyle\sum\limits_{z=1}^{N}}
\left[  \beta J_{1}\sigma_{z}\sigma_{z+1}+\beta J_{2}\sigma_{z}\sigma
_{z+2}\right]  +%
{\displaystyle\sum\limits_{z=1}^{N}}
\beta\eta_{z}\sigma_{z}\right]  ,
\end{equation}
where $\sigma_{z}$ is a short-hand notation for $\sigma_{x,y,z}$. We then have%
\begin{equation}
\Phi=-\frac{1}{\beta}N^{2}\ln Z_{01}-N^{2}%
{\displaystyle\sum\limits_{z=1}^{N}}
J_{0}m_{z}^{2}+N^{2}%
{\displaystyle\sum\limits_{z=1}^{N}}
\eta_{z}m_{z},
\end{equation}
where%
\begin{equation}
m_{z}=\left\langle \sigma_{x,y,z}\right\rangle _{0}=\frac{1}{\beta}%
\frac{\partial}{\partial\eta_{z}}\ln Z_{01}.\label{mz}%
\end{equation}
Note that $m_{z}$ depends on the set of field variables $\left\{  \eta
_{z}\right\}  $. In other words, $m_{z}=m_{z}\left(  \left\{  \eta
_{z}\right\}  \right)  $. The minimization of $\Phi$ leads to the condition
$\eta_{z}=J_{0}m_{z}$, which should be inserted into Eq. (\ref{mz}) to produce
a set of self-consistent equations for $\left\{  m_{z}\right\}  $. With the
trivial correspondence $J_{0}\rightarrow J_{0}/4$ to account for the
four-coordination of the spins on the $x-y$ planes, these expressions lead to
the same results already obtained in this Section for the LC-ANNNI model.

\section{Analysis of the numerical results}

\subsection{Paramagnetic critical lines}

It is easy to obtain an expression for the transition lines separating the
paramagnetic and the ordered phases. Consider an expansion of $\phi$, given by
Eq. (\ref{fi0}), in terms of the effective magnetizations,%
\begin{equation}
\phi=\frac{J_{0}}{2N}\sum_{z=1}^{N}m_{z}^{2}-\frac{\beta J_{0}^{2}}{2N}%
\sum_{z,z^{\prime}}\left\langle \sigma_{z}\sigma_{z^{\prime}}\right\rangle
_{0}m_{z}m_{z^{\prime}}+...,
\end{equation}
where $\left\langle \sigma_{z}\sigma_{z^{\prime}}\right\rangle _{0}$ is a
zero-field pair correlation of an Ising chain,%
\begin{equation}
\left\langle \sigma_{z}\sigma_{z^{\prime}}\right\rangle _{0}=\frac{1}%
{Z_{I}(\left\{  0\right\})  }\sum_{\left\{  \sigma_{z}\right\}  }\sigma
_{z}\sigma_{z^{\prime}}\exp\left[  \sum_{z=1}^{N}\left(  \beta J_{1}\sigma
_{z}\sigma_{z+1}+\beta J_{2}\sigma_{z}\sigma_{z+2}\right)  \right]  ,
\end{equation}
which has been calculated by Stephenson \cite{stephenson70}.

The paramagnetic transition lines come from%
\begin{equation}
1=\beta J_{0}\chi_{0}\left(  q_{\max}\right)  ,\label{para}%
\end{equation}
where%
\begin{equation}
\chi_{0}\left(  q\right)  =\sum_{h}\left\langle \sigma_{z}\sigma
_{z+h}\right\rangle _{0}\exp\left(  -iqh\right)  =\frac{A+B\cos q}{C+D\cos
q+E\cos^{2}q},
\end{equation}
and the coefficients $A$ to $E$ are real functions of $\beta J_{1}$ and $\beta
J_{2}$, which have been explicitly obtained by Stephenson \cite{stephenson77}%
\cite{harada83}. We can use these expressions, with $J_{0}=4J_{1}=4J>0$, to
draw the paramagnetic lines, given by Eq. (\ref{para}), and to locate the
Lifshitz point in a phase diagram in terms of $k_{B}T/J$ and the parameter
$p$, and then compare with the well-known results from the old layer-by-layer
mean-field approximations. Due to the\ inclusion of extra fluctuations along
the axial direction, it is not surprising that the critical temperature is
slightly smaller than the old mean-field values. Similar results had already
been obtained in previous calculations for the ANNNI model, which were,
however, limited to the analysis of the paramagnetic border \cite{pires82}%
\cite{harada83}.

\subsection{Phase diagrams}

The mean-field equations (\ref{eme}) were solved numerically using quadruple
precision. All integer values of $n$ should be considered to obtain the
modulated structures that minimize the free energy functional (\ref{fi2}).
However, since this is not feasible, earlier calculations were limited to a
relatively small set of modulated phases (with $n\lesssim20$) \cite{bakb80}%
\cite{yokoi81}. A real advance in these calculations has been achieved by
Selke and Duxbury \cite{selked84}\cite{selke88}, with the proposal of a
structure combination branching mechanism to explain the onset of different
commensurate phases at increasing temperatures. According to this mechanism,
at $T>0$, the boundaries between two adjacent modulated phases, which we call
$A$ and $B$, will become unstable against a new intervening phase $AB$. For
example, using the standard notation for the modulated phases of the ANNNI
model \cite{selke88}, consider the ordered structures (i) $A=\left\langle
3\right\rangle $, which consists of three planes of (predominantly) $+$ spins
followed by three planes of $-$ spins, in a periodic pattern along the axial
direction, and (ii) $B=\left\langle 32\right\rangle $, which consists of three
planes of $+$ spins, followed by two planes of $-$ spins. These phases will
become unstable against the new intervening phase $AB=\left\langle
332\right\rangle $. In more general terms, given the phases $A=\left\langle
32^{j-1}\right\rangle $ and $B=\left\langle 32^{j}\right\rangle $, with $j=1$,
$2$, $3$, ..., \ we have the new intervening phase $AB=\langle32^{j-1}%
32^{j}\rangle$.

In Fig. 1, we show the main commensurate structures in the $T-p$ phase
diagram, with $J_{0}=J_{1}=J>0$, and $p=-J_{2}/J$. We have four large regions:
paramagnetic, ferromagnetic (ferro), the antiphase $\left\langle
2\right\rangle $, and the large region of modulated structures. According to
the expectations, the paramagnetic critical border meets tangentially the
first-order ferro-modulated border at the Lifshitz point (LP). As the
temperature increases, we checked that long-period structures become stable in
smaller regions of this phase diagram. We recall that the notation
$\left\langle 3\right\rangle $ means that there are $3$ planes of $+$ spins
followed by $3$ planes of $-$ spins along the axial direction. Although we use
a different temperature scale, the general qualitative topology of this phase
diagram is the same as obtained in the earlier mean-field calculations.
Modulated phases $\left\langle 2\right\rangle $, $\left\langle 3\right\rangle
$, $\left\langle 4\right\rangle $, and $\left\langle 32\right\rangle $ still
occupy large portions of the modulated region.

In the vicinity of the multiphase point, both the LC-ANNNI and the standard
ANNNI model display the same qualitative features. Simple periodic structures
of the type $\left\langle 32^{j}\right\rangle $ still play a major role at low
temperatures. In addition, the $\left\langle 4\right\rangle $ phase displays a
range of stability between $\left\langle \infty\right\rangle $ and
$\left\langle 3\right\rangle $. This behavior is consistent with the
predictions of the domain-wall analysis for sufficiently anisotropic cases
\cite{szpilkaf86}.%


In all of our calculations, we have fully confirmed the branching mechanism,
which keeps working in the modulated regions, as the temperature increases,
and helps to drastically reduce the number of phases to be analyzed. The wave
number of the new modulated phase is in the interval between the wave numbers
of the parent modulated phases, according to the rule for the construction of
a Farey tree. As the temperature increases, the wave numbers of the new
modulated phases, in units of $2\pi$, will tend to cover all the rational numbers.%
\begin{figure}
[!htb]
\begin{center}
\includegraphics[
scale=0.45
]%
{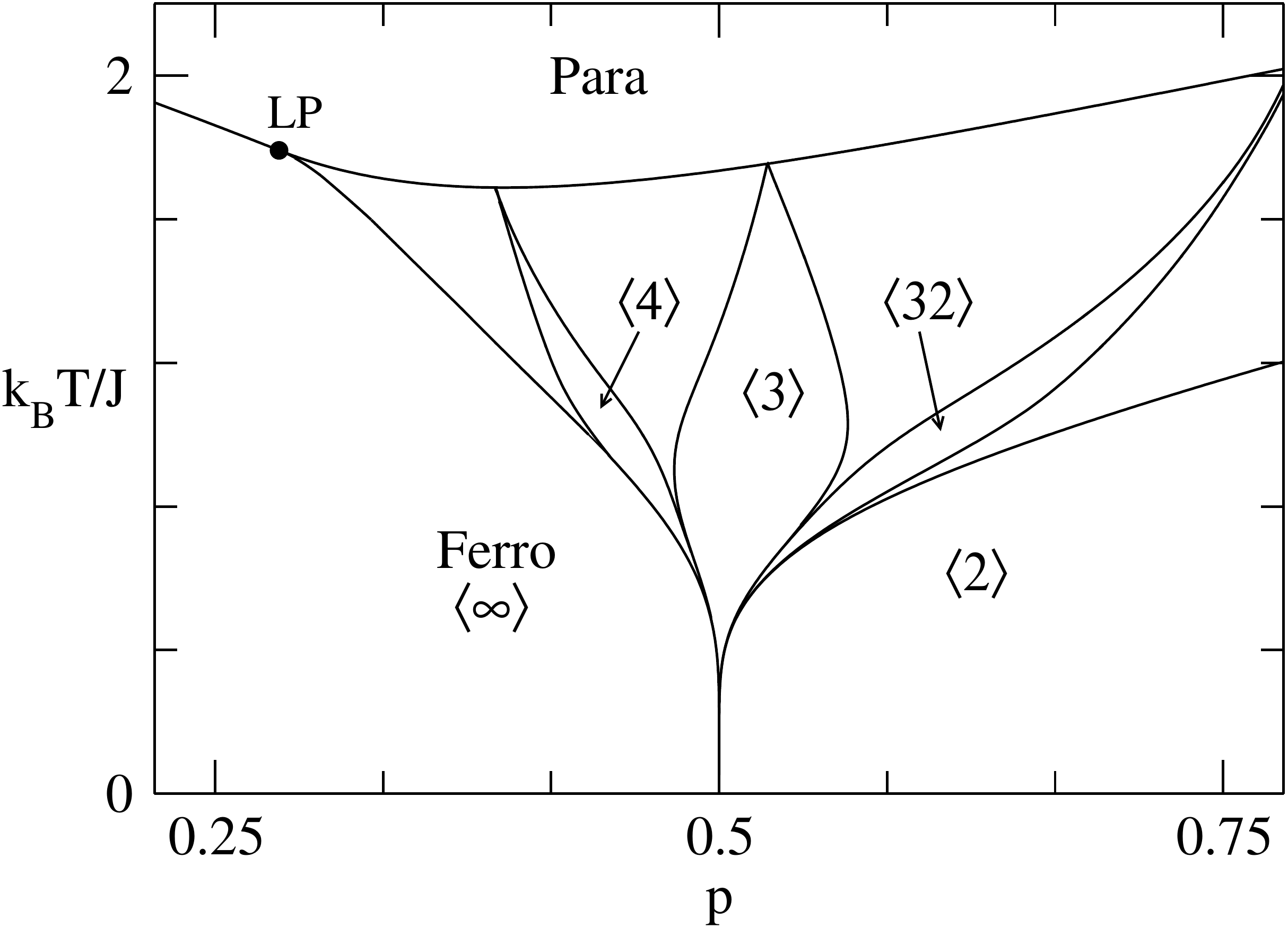}%
\caption{Main commensurate phases in the global $T-p$ phase diagram of the
LC-ANNNI model (with $J_{0}=J_{1}=J>0$, $p=-J_{2}/J_{1}>0$). The paramagnetic
border comes form Eq.(\ref{para}). The Lifshitz point is located at
$p_{LP}=0.28172...$ and $k_{B}T_{LP}/J=1.79152...$}%
\end{center}
\end{figure}

%

In Fig.2, we draw a typical graph of the wave number of modulated phases
versus temperature for a particular value of the parameter of competition,
$p=0.57$ (and with $J_{0}=J_{1}=J>0$). We have used the branching mechanism,
with quadruple precision, to draw this graph (and the graphs of the following
figures). The simple periodic structures are associated with wide plateaus of
stability, and there are higher-order commensurate phases between the main
commensurate structures.
\begin{figure}
[!htb]
\begin{center}
\includegraphics[
scale=0.45
]%
{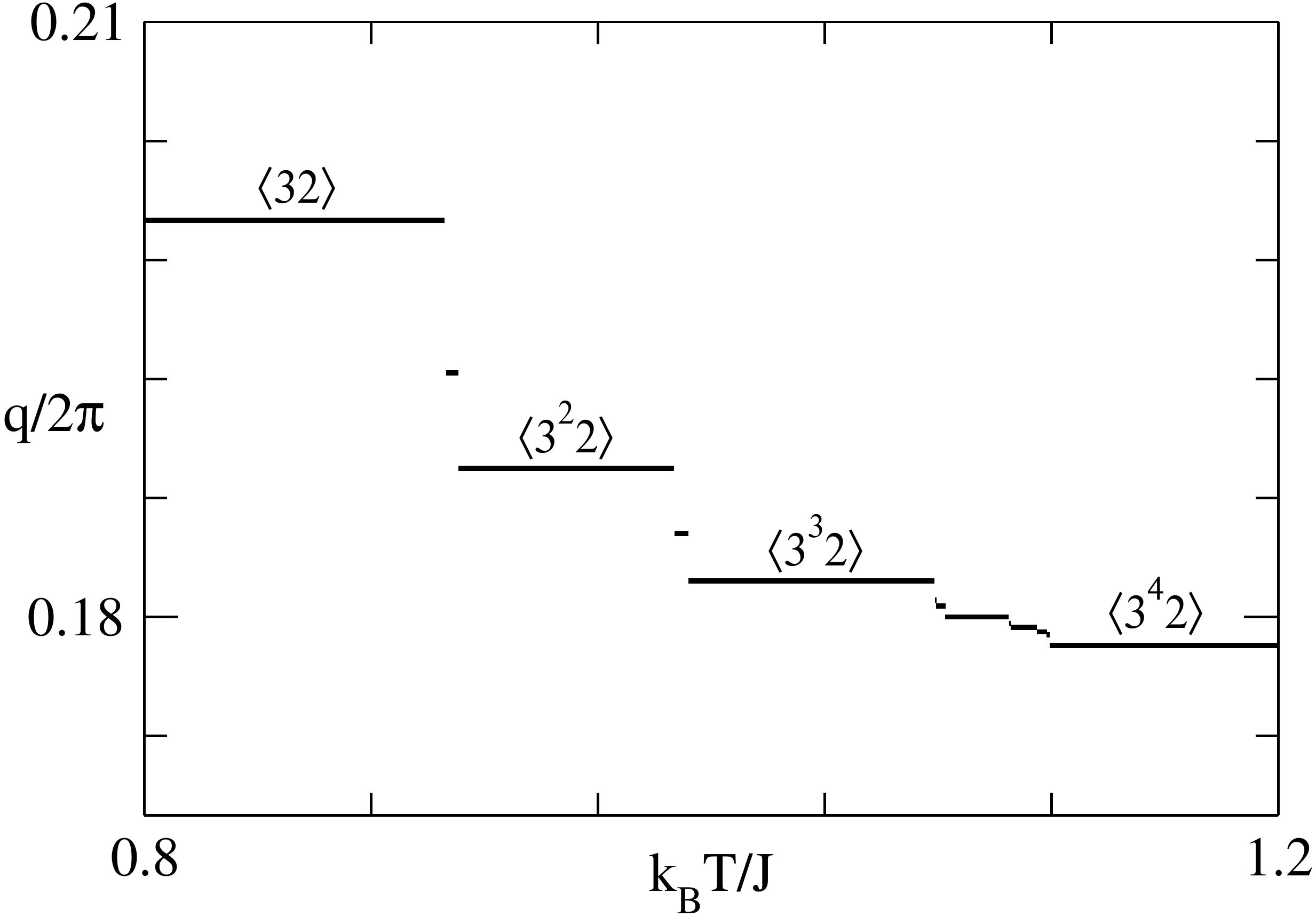}%
\caption{Typical graph of the wave number of the main modulated phases versus
temperature for $p=0.57$ (with $J_{0}=J_{1}=J>0$). Note the existence of
higher-order commensurate phases in the intervals between the plateaus.}%
\end{center}
\end{figure}

In the immediate vicinity of the Lifshitz point, we can analytically show that
the ferro-modulated border is discontinuous. In order to further check the
nature of this border, in Fig. 3 we draw a graph of the main wave number of
the modulated structures as a function of $p$, for fixed temperature
$k_{B}T/J=1.1$ (with $J_{0}=J_{1}=J>0$). We indicate the modulated phases
associated with the largest plateaus, $\left\langle 3\right\rangle $,
$\left\langle 43\right\rangle $, and $\left\langle 4\right\rangle $, and draw
this graph to point out the discontinuous character of the transition to the
ferromagnetic phase ($q=0$), which does agree with the old mean-field and
domain-wall calculations (and disagrees with the variational calculations of
Gendiar and Nishino \cite{gendiarn05}).%

%
%
\begin{figure}
[!htb]
\begin{center}
\includegraphics[
scale=0.45
]%
{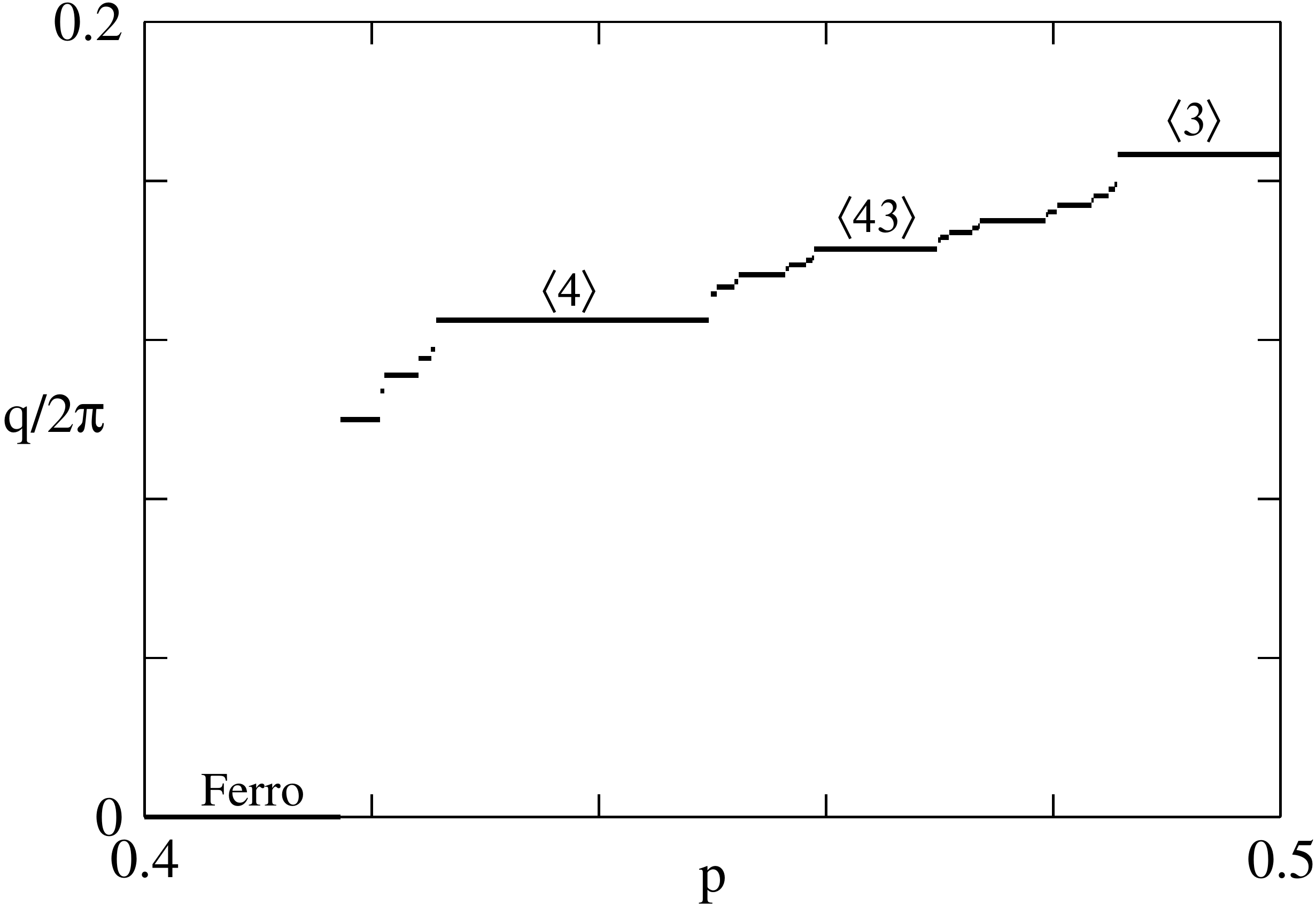}%
\caption{Graph of the main wave number of the modulated structures as a
function of $p$, for fixed temperature $k_{B}T/J=1.1$ (with $J_{0}=J_{1}%
=J>0$).}%
\end{center}
\end{figure}

\subsection{Devil%
\'{}%
s staircases}

At fixed temperature, as we change the parameter $p$, the wave number locks in
rational values, which gives rise to a sequence of phase transitions. At
intermediate temperatures, many distinct commensurate phases are locked in
finite regions of stability. For example, the simple periodic phases
$\langle3^{n}2\rangle$ and $\langle32^{n}\rangle$ lock in large intervals of
the parameter $p$ (see Fig. 4). These results are in quantitative disagreement
with the claims of some recent Monte Carlo simulations for the ANNNI model,
which seem to support much narrower ranges of stability of the modulated
structures \cite{zhangc10}. In contrast, phases $\langle3\rangle$ and
$\langle32\rangle$ are shown to occupy small regions in the phase diagram, as
it has been obtained in these Monte Carlo simulations.

We now turn to the question of the fractal dimension of the $q$ versus $p$
graphs. We use a well-known box-counting algorithm to estimate the fractal
dimension of the set of points that remain in an interval of values of $q$ if
we subtract all of the intervals corresponding to plateaux of the commensurate
phases larger than a certain (limiting small) width. For example, consider the
plateaux in the graph of $q$ versus $p$ of Fig. 4, and look at the interval
between $q_{i}=0.5$ and $q_{f}=0.8$. Calculate the difference $X\left(
\epsilon\right)  $ between the width $q_{f}-q_{i}$ and the sum of the
intervals corresponding to the commensurate phases with plateaux of widths
larger than a certain length $\epsilon>0$. The slope of a log-log plot of
$X\left(  \epsilon\right)  /\epsilon$ versus $1/\epsilon$, in the limit of
small values of $\epsilon$, gives the Hausdorff fractal dimension $D_{0}$
associated with the (much smaller) plateaux of the remaining set of phases. If
$D_{0}<1$, the remaining (presumably incommensurate) phases occupy a fractal
set (of zero measure). This is the anticipated situation at intermediate
temperatures, at least not so close to the paramagnetic transition. If
$D_{0}<1$, we say that we have a complete devil%
\'{}%
s staircase.%
\begin{figure}
[!htb]
\begin{center}
\includegraphics[
scale=0.45
]%
{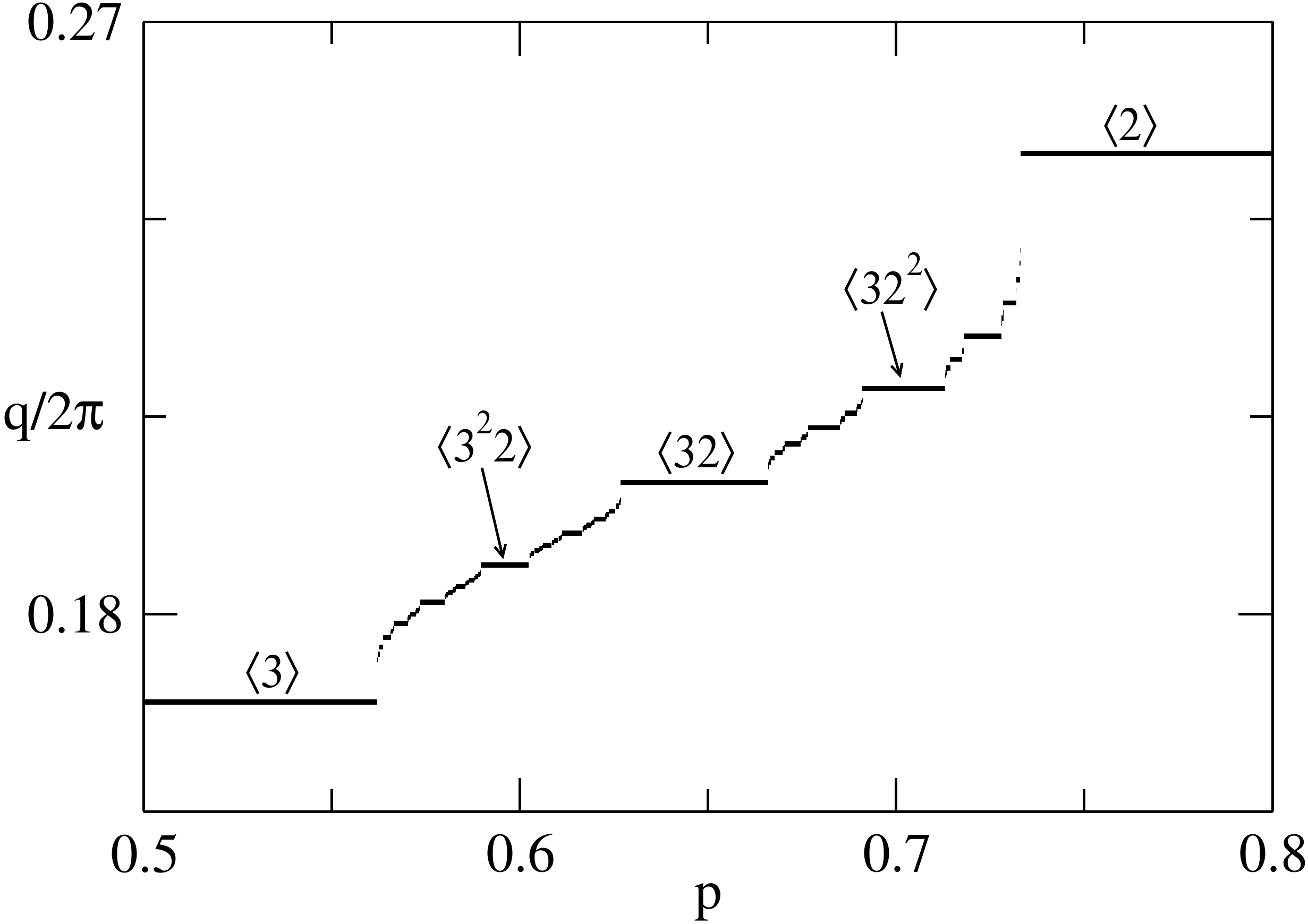}%
\caption{Wave number versus $p$ for $k_{B}T/J=1.125$. There is a cascade of
phase transitions as the wave number $q$ goes from $2\pi/6$ (in the
$\langle3\rangle$ phase) to $2\pi/4$ (in the $\langle2\rangle$ phase). Many
distinct commensurate modulated phases are stable. We show phases associated
with plateaus of width $\Delta p>10^{-5}$ only.}%
\end{center}
\end{figure}


In Fig. 5, we show numerically obtained plots of $\log\left[  X\left(
\epsilon\right)  /\epsilon\right]  $ versus $\log\left[  1/\epsilon\right]  $
for $k_{B}T/J=1.1$ (graph A) and $k_{B}T/J=1.15$ (graph B). From these
straight lines, we obtain the Hausdorff dimensions, which are plotted in Fig.
6, for a few increasing values of temperature, below the paramagnetic critical
line. As $D_{0}<1$, we have complete devil%
\'{}%
s staircases (in other words, at these temperatures, incommensurate phases
occupy a region of fractal measure). Since these calculations are much harder
at higher temperatures, we can only claim that we have numerical evidence that
$D_{0}<1$ increases with temperature, and seems to reach a value smaller than
$D_{0}\approx0.8$, which is in agreement with numerical calculations for
several mapping problems \cite{biham89}.
\begin{figure}
[!htb]
\begin{center}
\includegraphics[
scale=0.45
]%
{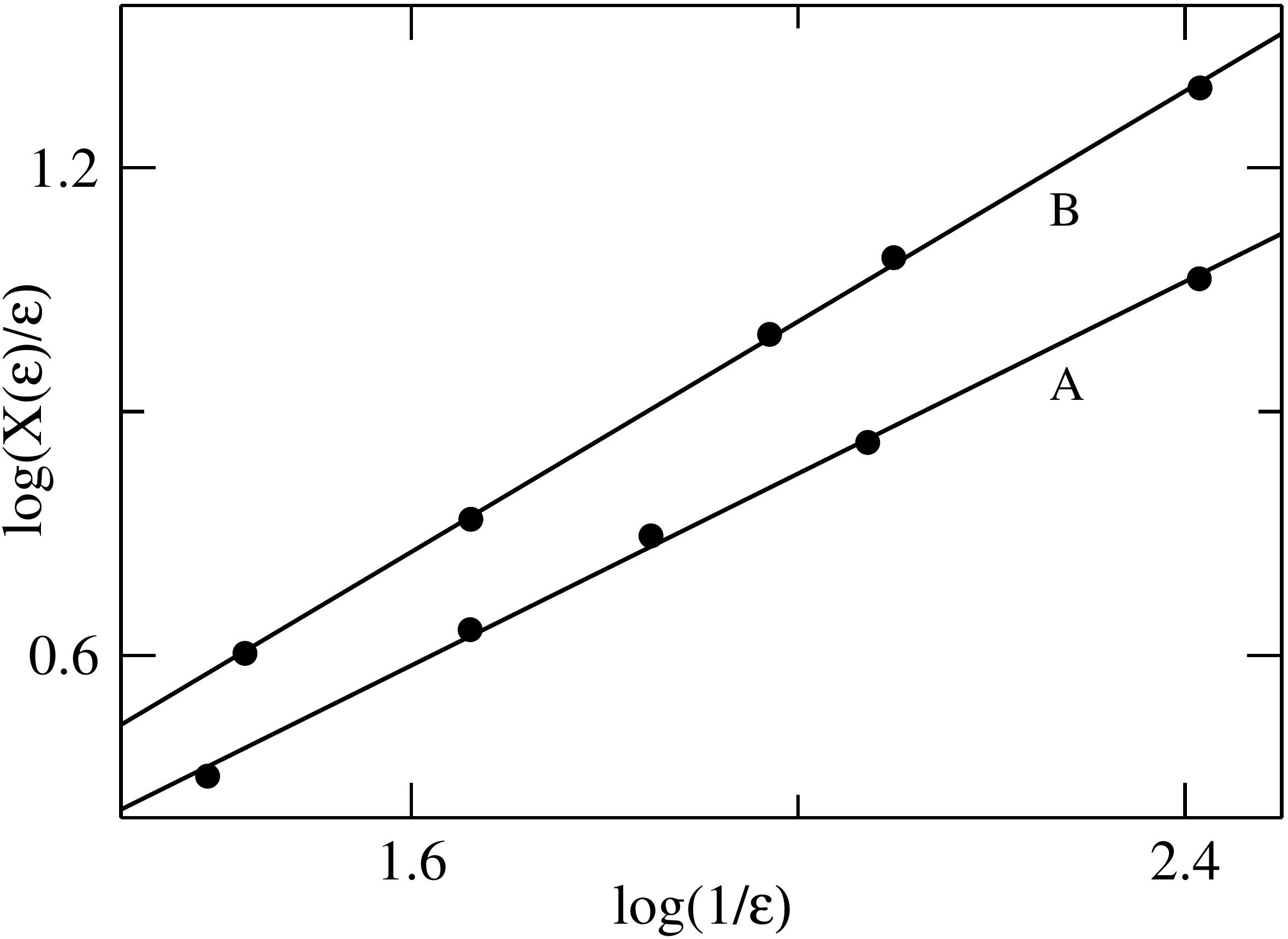}%
\caption{Plots of $\log\left[  X\left(  \epsilon\right)  /\epsilon\right]  $
versus $\log\left[  1/\epsilon\right]  $ for $k_{B}T/J=1.1$ (graph A) and
$k_{B}T/J=1.15$ (graph B).}%
\end{center}
\end{figure}
\begin{figure}
[!htb]
\begin{center}
\includegraphics[
scale=0.45
]%
{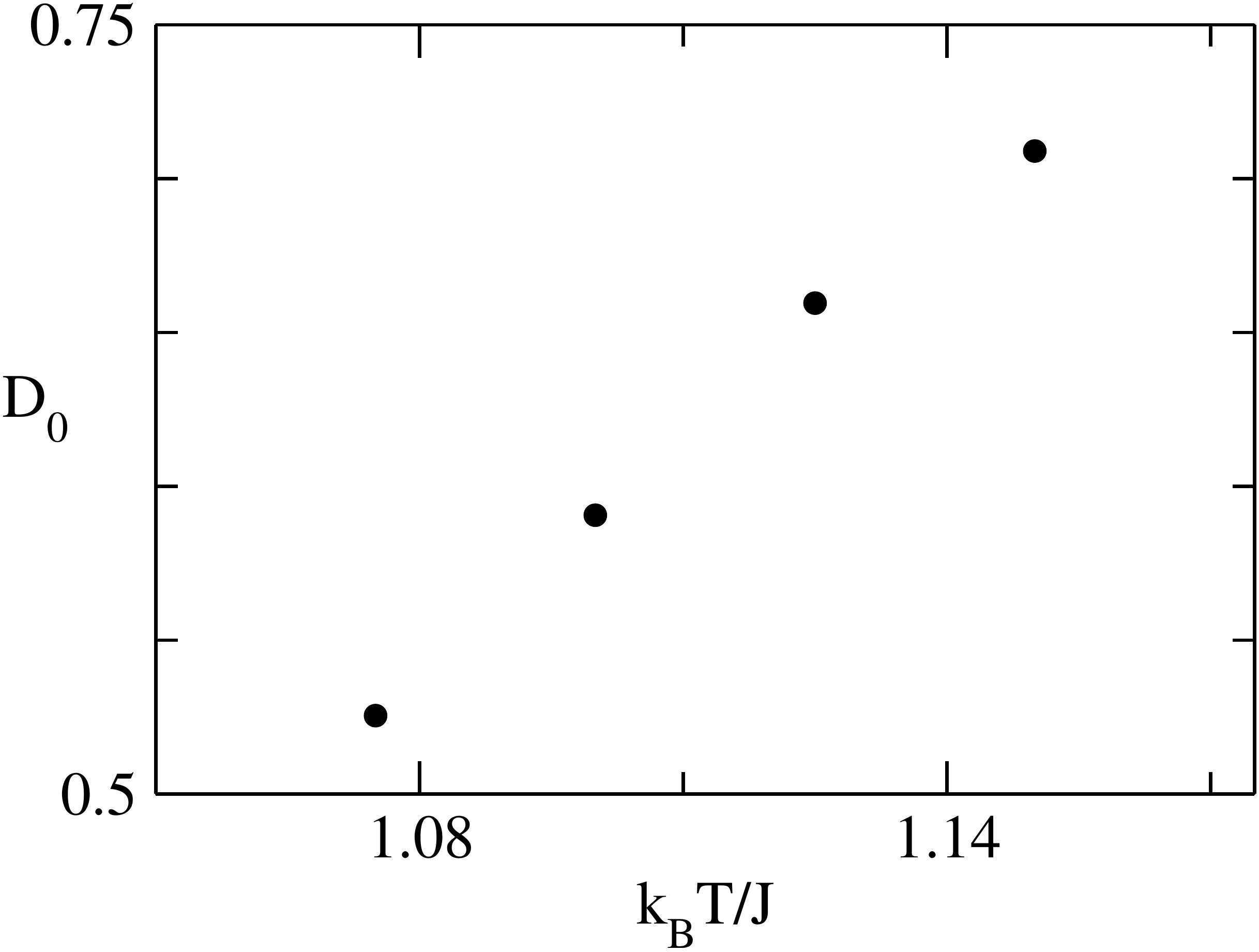}%
\caption{Hausdorff fractal dimension $D_{0}$ as function of temperature
($k_{B}T/J$, with $J_{0}=J$). Note that $D_{0}<1$ increases wih
temperature.\bigskip}%
\end{center}
\end{figure}

\section{Conclusions}

We investigated the equilibrium behavior of a spin-$1/2$ Ising system with
axial competing interactions between nearest and next-nearest neighbors, and
infinite-range interactions between spins on the sites of planes perpendicular
to the axial direction. This system may be regarded as obtained from a
particular limit of infinite coordination of the layers of the ANNNI model on
a\ hypercubic lattice. The same results can also be obtained from a mean-field
variational treatment of the ANNNI model on a hypecubic lattice \cite{yokoi81}%
, if we use a trial Hamiltonian formed by set of independent Ising chains,
with next-nearest-neighbor interactions, in a position-dependent field.

On the basis of an expression for the free energy, we perform numerical
calculations to check the main features of the phase diagram and the main
spatially modulated phases. At low temperatures, in accordance with the
domain-wall analysis of the ANNNI model, the ferromagnetic and the
$\langle2\rangle$ phases melt via a first-order transition into the modulated
phases. At quite low temperatures, the $\langle4\rangle$ phase has a small
range of stability, between the ferromagnetic and $\langle3\rangle$ phases,
which depends upon the model parameters. Higher-order commensurate phases
become stable with increasing temperatures. We confirm the existence of a
branching mechanism to stabilize long-period modulated structures at
increasing temperatures (note that we are assuming $J_{0}=J_{1}$). In terms of
either temperature or the parameter of competition, the main wave number of
the modulated phases divided by $2\pi$ locks at rational values. We draw some
graphs of these wave numbers as a function of $p$, for fixed values of $T$,
and perform a detailed numerical analysis of the fractal character of the
associated devil%
\'{}%
s staircases. We calculate the Hausdorff dimension $D_{0}<1$ of these fractal
structures, and show that $D_{0}$ increases with temperature, with a limiting
value $D_{0}\approx0.8$, which seems to be a common feature of several
problems represented by area-preserving maps. We support the picture that
simple periodic phases play the main role in the ordered region of the $T-p$
phase diagram.

\end{document}